\documentclass[aps]{revtex4}%
\usepackage{amsfonts}
\usepackage{amsmath}
\usepackage{amssymb}
\usepackage{graphicx}%
\begin{document}
\title[Annular well in graphene]{Electrons scattering in the monolayer graphene with a band-asymmetric annular
potential well }
\author{Natalie E. Firsova}
\affiliation{Institute for Problems of Mechanical Engineering, the Russian Academy of
Sciences, St. Petersburg 199178, Russia}
\author{Sergey A. Ktitorov}
\affiliation{A.F. Ioffe Physical-Technical Institute, the Russian Academy of Sciences,
Polytechnicheskaya str. 26, St. Petersburg 194021, Russia}
\keywords{Dirac equation, S-matrix, T-matrix}

\begin{abstract}
Electron scattering in the monolayer graphene with short-range impurities
modelled by the annular well with a band-asymmetric potential has been
considered. Band-asymmetry of the potential resulted in the mass (gap)
perturbation in the Dirac equation. Exact explicit formulae for the scattering
matrix have been derived. The results are presented in terms of the scattering
phases and in the geometrical form of a relation between some 2-vectors. The
characteristic equation is obtained. It has a form of the orthogonality
condition. An approximate calculation of observables in terms the scattering
theory results is outlined.

\end{abstract}
\maketitle









\section{Introduction}

We consider in this work the electrons scattering in the 2+1 Dirac equation
model of the monolayer graphene due to the short-range perturbations.
Short-range potential impurities in graphene were considered in works
\cite{basko}, \cite{novikov}, \cite{matulis}. In our work \cite{we}, a new
model of the short-range impurities in graphene was considered taking into
account the obvious fact that the Kohn-Luttinger matrix elements of the
short-range perturbation calculated on the upper and lower band wave functions
are not equal in a general case. This means that the perturbation must be
generically described by a Hermitian matrix. The diagonal matrix case
corresponding to a presence of the potential and mass perturbation was
considered. Annular geometry of the perturbation allows us to exclude a
perturbation at small distances, which are unphysical in a crystal. Delta
function model of the short-range diagonal matrix perturbation was considered
in our works \cite{we}, \cite{we2}. We study here the potential and mass well
model. We consider scattering of electrons by this potential and obtain exact
explicit formulae for the scattering data. We also discuss application of
these formulae for calculation of observables.

\section{Basic equations}

The Dirac equation for electronic states in graphene for our model of matrix
annular well described above reads
\begin{equation}
\left(  -i\hbar v_{F}\sum_{\mu=1}^{2}\alpha_{\mu}\partial_{\mu}-\beta\left(
m+\delta m\left(  r\right)  \right)  v_{F}^{2}{}\right)  \psi=\left(
E-V\left(  r\right)  \right)  \psi,\label{diracgeneral}%
\end{equation}
where $v_{F}$ is the Fermi velocity of the band electrons, $\alpha_{\mu}$
$\beta$ are the Dirac matrices $\beta=\sigma_{3},$ $\alpha_{1}=\sigma_{1},$
$\alpha_{2}=i\sigma_{2},$ $\sigma_{i}$ are the Pauli matrices, $2mv_{F}{}%
^{2}=E_{g}$ is the electronic bandgap, $\psi\left(  \mathbf{r}\right)  $ is
the two-component spinor. The electronic gap can appear in the graphene
monatomic film lying on the substrate because of the sublattices mutual shift
\cite{gap}. The spinor structure takes into account the two-sublattice
configuration of graphene.$\ \delta m\left(  \mathbf{r}\right)  $ and
$V(\mathbf{r})$ are the local perturbations of the mass (gap) and the chemical potential:%

\begin{equation}
V\left(  r\right)  =-a\Delta\left(  r\right)  ,\text{ \ \ }\delta m\left(
r\right)  =-b\Delta\left(  r\right)  , \label{perturb}%
\end{equation}

where $\Delta\left(  r\right)  $ is determined as follows%
\begin{equation}
\Delta\left(  r\right)  =\left\{
\begin{array}
[c]{c}%
0,\text{ \ \ }0\leq r<r_{1}\text{ \ \ (region I),}\\
\\
1,\text{ \ \ }r_{1}\leq r<r_{2}\text{ \ \ (region II),}\\
\\
0,\text{ \ \ }r\geq r_{2}\text{ \ \ (region III).}%
\end{array}
\right.  . \label{delta}%
\end{equation}

\bigskip Let us present the two-component spinor in the form%
\begin{equation}
\psi_{j}(\mathbf{r},t)=\frac{\exp\left(  -iEt\right)  }{\sqrt{r}}\left(
\begin{array}
[c]{c}%
f_{j}\left(  r\right)  \exp\left[  i\left(  j-1/2\right)  \varphi\right] \\
\\
g_{j}\left(  r\right)  \exp\left[  i\left(  j+1/2\right)  \varphi\right]
\end{array}
\right)  , \label{spinor}%
\end{equation}
where $j$ is the pseudospin quantum number; $j=\pm1/2,$ $\pm3/2,\ldots$. In
opposite to the relativistic theory, this quantum number has nothing to do
with the real spin and indicates a degeneracy in the biconic Dirac point. The
upper $f_{j}\left(  r\right)  $ and lower $g_{j}\left(  r\right)  $ components
of the spinor satisfy the equations set
\begin{equation}
\frac{dg_{j}}{dr}+\frac{j}{r}g_{j}-\left(  E-m\right)  f_{j}=\left(
a+b\right)  \Delta\left(  r\right)  f_{j}, \label{componenteq1}%
\end{equation}

\begin{equation}
-\frac{df_{j}}{dr}+\frac{j}{r}f_{j}-\left(  E+m\right)  g_{j}=\left(
a-b\right)  \Delta\left(  r\right)  g_{j}. \label{componeq2}%
\end{equation}

These equations have a symmetry:
\begin{equation}
f_{j}\leftrightarrow g_{j},\text{ }E\rightarrow-E,\text{ }j\rightarrow
-j,\text{ }a\rightarrow-a. \label{symm}%
\end{equation}

\section{Scattering matrix and characteristic equation}

When $r\overline{\in}\left[  r_{1},r_{2}\right)  $ (in regions I and III), we
obtain from eq. (\ref{componenteq1}) and eq. (\ref{componeq2}):

\begin{equation}
\frac{d^{2}f_{j}}{dr^{2}}+\left[  E^{2}-m^{2}-\frac{j\left(  j-1\right)
}{r^{2}}\right]  f_{j}=0. \label{secondorder}%
\end{equation}
This equation is related to the Bessel one. Its solution in the region I
reads:%
\begin{align}
f_{j}\left(  r\right)   &  =C_{1}\sqrt{\kappa r}J_{j-1/2}\left(  \kappa
r\right)  ,\label{region1f}\\
g_{j}\left(  r\right)   &  =C_{1}\sqrt{\frac{E-m}{E+m}}\sqrt{\kappa
r}J_{j+1/2}\left(  \kappa r\right)  , \label{region1g}%
\end{align}
where $\kappa=\sqrt{E^{2}-m^{2}},$ $J_{n}\left(  x\right)  $ is the Bessel function.

\qquad Let us introduce the function $\varphi_{j}\left(  r\right)  :$%
\begin{equation}
\varphi_{j}\left(  r\right)  \equiv f_{j}/g_{j}. \label{phi}%
\end{equation}

We obtain from (\ref{region1f}) and eq. (\ref{region1g}) for the region $0\leq
r<r_{1}$:%
\begin{equation}
\varphi_{j}^{I}\left(  \kappa r\right)  =\sqrt{\frac{E+m}{E-m}}\frac
{J_{j-1/2}\left(  \kappa r\right)  }{J_{j+1/2}\left(  \kappa r\right)  }
\label{phi1}%
\end{equation}
We have from (\ref{delta}), (\ref{componenteq1}), and (\ref{componeq2}) within
the region II:%
\begin{equation}
\frac{dg_{j}}{dr}+\frac{j}{r}g_{j}-\left(  E-m\right)  f_{j}=\left(
a+b\right)  f_{j}, \label{eq2g}%
\end{equation}

\begin{equation}
-\frac{df_{j}}{dr}+\frac{j}{r}f_{j}-\left(  E+m\right)  f_{j}=\left(
a-b\right)  g_{j}. \label{eq2f}%
\end{equation}
These equations can be re-written in the form
\begin{equation}
\frac{dg_{j}}{dr}+\frac{j}{r}g_{j}-\left(  \widetilde{E}-\widetilde{m}\right)
f_{j}=0, \label{componenteq1a}%
\end{equation}%
\begin{equation}
-\frac{df_{j}}{dr}+\frac{j}{r}f_{j}-\left(  \widetilde{E}+\widetilde
{m}\right)  g_{j}=0, \label{componenteq2a}%
\end{equation}
where $\widetilde{E}=E+a,$ $\widetilde{m}=m-b.$ Thus we have the following
second-order equation in the region $r_{1}\leq r<r_{2}:$%

\begin{equation}
\frac{d^{2}f_{j}}{dr^{2}}+\left[  \widetilde{E}^{2}-\widetilde{m}^{2}%
-\frac{j\left(  j-1\right)  }{r^{2}}\right]  f_{j}=0. \label{secondorder2}%
\end{equation}

Then the function $\varphi_{j}\left(  r\right)  $ (\ref{phi}) can be written
in the region II as follows:%
\begin{equation}
\varphi_{j}^{II}\left(  r\right)  =\sqrt{\frac{\widetilde{E}+\widetilde{m}%
}{\widetilde{E}-\widetilde{m}}}\frac{J_{j-1/2}\left(  \widetilde{\kappa
}r\right)  +C_{j}N_{j-1/2}\left(  \widetilde{\kappa}r\right)  }{J_{j+1/2}%
\left(  \widetilde{\kappa}r\right)  +C_{j}N_{j+1/2}\left(  \widetilde{\kappa
}r\right)  }, \label{phi2}%
\end{equation}

where $\widetilde{\kappa}^{2}=\widetilde{E}^{2}-\widetilde{m}.$ Similarly we
obtain for the region III:%
\begin{equation}
\varphi_{j}^{III}\left(  r\right)  =\sqrt{\frac{E+m}{E-m}}\frac{H_{j-1/2}%
^{\left(  2\right)  }(\kappa r)+S_{j}H_{j-1/2}^{\left(  1\right)  }(\kappa
r)}{H_{j+1/2}^{\left(  2\right)  }(\kappa r)+S_{j}H_{j+1/2}^{\left(  1\right)
}(\kappa r)} \label{phi3}%
\end{equation}
Continuity of the spinor components leads to matching conditions for the
functions $\varphi_{j}\left(  r\right)  $ at the boundaries between regions I,
II and III. We obtain in result the following expressions for the coefficients
$C_{j}$ and $S_{j}:$%
\begin{equation}
C_{j}=-\frac{\sqrt{\frac{E+m}{E-m}}J_{j-1/2}\left(  \kappa r_{1}\right)
J_{j+1/2}\left(  \widetilde{\kappa}r_{1}\right)  -\sqrt{\frac{\widetilde
{E}+\widetilde{m}}{\widetilde{E}-\widetilde{m}}}J_{j+1/2}\left(  \kappa
r_{1}\right)  J_{j-1/2}\left(  \widetilde{\kappa}r_{1}\right)  }{\sqrt
{\frac{\widetilde{E}+\widetilde{m}}{\widetilde{E}-\widetilde{m}}}%
N_{j-1/2}\left(  \widetilde{\kappa}r_{1}\right)  J_{j+1/2}\left(  \kappa
r_{1}\right)  -\sqrt{\frac{E+m}{E-m}}N_{j+1/2}\left(  \widetilde{\kappa}%
r_{1}\right)  J_{j-1/2}\left(  \kappa r_{1}\right)  }, \label{C}%
\end{equation}

\begin{equation}
S_{j}=-\frac{F_{j}^{\left(  2\right)  }}{F_{j}^{\left(  1\right)  }},
\label{S}%
\end{equation}

where
\begin{align}
F^{\left(  \alpha\right)  }  &  =\sqrt{\frac{E+m}{E-m}}H_{j-1/2}^{\left(
\alpha\right)  }(\kappa r_{2})\left[  J_{j+1/2}\left(  \widetilde{\kappa}%
r_{2}\right)  +C_{j}N_{j+1/2}\left(  \widetilde{\kappa}r_{2}\right)  \right]
-\nonumber\\
&  -\sqrt{\frac{\widetilde{E}+\widetilde{m}}{\widetilde{E}-\widetilde{m}}%
}H_{j+1/2}^{\left(  \alpha\right)  }(\kappa r_{2})\left[  J_{j-1/2}\left(
\widetilde{\kappa}r_{2}\right)  +C_{j}N_{j-1/2}\left(  \widetilde{\kappa}%
r_{2}\right)  \right]  \label{F}%
\end{align}

The constant $C_{j}$ is determined by the formula (\ref{C}). Here $\alpha$
takes values $0,1.$ The constant $S_{j}$ is a phase factor of the outgoing
wave, i. e. the S-matrix element in the angular momentum representation. Since
$H_{n}^{\left(  2\right)  }\left(  z\right)  =H_{n}^{\left(  1\right)  \ast
}\left(  z\right)  $ for real $z,$ the scattering matrix is unitary everywhere
on the continuous spectrum. Equations\ (\ref{C}), (\ref{S}) and (\ref{F})
solve the electron scattering problem for the given potential. The denominator
of $S_{j}(E)$ is just the left-hand side of the characteristic equation for
the bound and resonance electronic states:%
\begin{equation}
F^{\left(  1\right)  }=0 \label{F1}%
\end{equation}

or%

\begin{equation}
\sqrt{\frac{E+m}{E-m}}H_{j-1/2}^{\left(  1\right)  }(\kappa r_{2})\left[
J_{j+1/2}\left(  \widetilde{\kappa}r_{2}\right)  +C_{j}N_{j+1/2}\left(
\widetilde{\kappa}r_{2}\right)  \right]  -\sqrt{\frac{\widetilde{E}%
+\widetilde{m}}{\widetilde{E}-\widetilde{m}}}H_{j+1/2}^{\left(  1\right)
}(\kappa r_{2})\left[  J_{j-1/2}\left(  \widetilde{\kappa}r_{2}\right)
+C_{j}N_{j-1/2}\left(  \widetilde{\kappa}r_{2}\right)  \right]  =0.
\label{char}%
\end{equation}

Excluding the constant $C_{j}$ from eq. (\ref{S}) and eq. $\left(
\text{\ref{char}}\right)  ,$ we obtain the explicit formula for $S_{j}$ and
the characteristic equation:%
\begin{align}
&  \frac{\sqrt{\frac{E+m}{E-m}}J_{j+1/2}\left(  \widetilde{\kappa}%
r_{1}\right)  J_{j-1/2}\left(  \kappa r_{1}\right)  -\sqrt{\frac{\widetilde
{E}+\widetilde{m}}{\widetilde{E}-\widetilde{m}}}J_{j-1/2}\left(
\widetilde{\kappa}r_{1}\right)  J_{j+1/2}\left(  \kappa r_{1}\right)  }%
{\sqrt{\frac{E+m}{E-m}}N_{j+1/2}\left(  \widetilde{\kappa}r_{1}\right)
J_{j-1/2}\left(  \kappa r_{1}\right)  -\sqrt{\frac{\widetilde{E}+\widetilde
{m}}{\widetilde{E}-\widetilde{m}}}N_{j-1/2}\left(  \widetilde{\kappa}%
r_{2}\right)  J_{j+1/2}\left(  \kappa r_{1}\right)  }\nonumber\\
&  =\frac{\sqrt{\frac{E+m}{E-m}}J_{j+1/2}\left(  \widetilde{\kappa}%
r_{2}\right)  H_{j-1/2}^{\left(  1\right)  }(\kappa r_{2})-\sqrt
{\frac{\widetilde{E}+\widetilde{m}}{\widetilde{E}-\widetilde{m}}}%
J_{j-1/2}\left(  \widetilde{\kappa}r_{2}\right)  H_{j+1/2}^{\left(  1\right)
}(\kappa r_{2})}{\sqrt{\frac{E+m}{E-m}}N_{j+1/2}\left(  \widetilde{\kappa
}r_{2}\right)  H_{j-1/2}^{\left(  1\right)  }(\kappa r_{2})-\sqrt
{\frac{\widetilde{E}+\widetilde{m}}{\widetilde{E}-\widetilde{m}}}%
N_{j-1/2}\left(  \widetilde{\kappa}r_{2}\right)  H_{j+1/2}^{\left(  1\right)
}(\kappa r_{2})}. \label{charfin}%
\end{align}

When $r_{1}=0$, $r_{2}=r_{0}$, we have a case of the simple round well; the
characteristic equation takes the form:%
\begin{equation}
\sqrt{\frac{\widetilde{E}+\widetilde{m}}{\widetilde{E}-\widetilde{m}}%
}J_{j-1/2}\left(  \widetilde{\kappa}r_{0}\right)  H_{j+1/2}^{\left(  1\right)
}(\kappa r_{0})=\sqrt{\frac{E+m}{E-m}}J_{j+1/2}\left(  \widetilde{\kappa}%
r_{0}\right)  H_{j-1/2}^{\left(  1\right)  }(\kappa r_{0}) \label{simple}%
\end{equation}

A geometric interpretation of these expressions for the S-matrix and
characteristic equation can be done here. Let us introduce the following
column vectors:%
\begin{equation}
\mathbf{J}_{j}\left(  r_{1}\right)  =\left(
\begin{array}
[c]{c}%
J_{j,1}\\
\\
J_{j,2}%
\end{array}
\right)  \equiv\left(
\begin{array}
[c]{c}%
\sqrt{\frac{\widetilde{E}+\widetilde{m}}{\widetilde{E}-\widetilde{m}}%
}J_{j+1/2}\left(  \kappa r_{1}\right) \\
\\
-\sqrt{\frac{E+m}{E-m}}J_{j-1/2}\left(  \kappa r_{1}\right)
\end{array}
\right)  , \label{J}%
\end{equation}

\begin{equation}
\mathbf{h}_{j}^{\left(  \alpha\right)  }\left(  r_{2}\right)  =\left(
\begin{array}
[c]{c}%
h_{j,1}\\
\\
h_{j,2}%
\end{array}
\right)  \equiv\left(
\begin{array}
[c]{c}%
\sqrt{\frac{\widetilde{E}+\widetilde{m}}{\widetilde{E}-\widetilde{m}}%
}H_{j+1/2}^{\left(  \alpha\right)  }(\kappa r_{2})\\
\\
-\sqrt{\frac{E+m}{E-m}}H_{j-1/2}^{\left(  \alpha\right)  }(\kappa r_{2})
\end{array}
\right)  \label{H}%
\end{equation}

and the matrix%
\begin{equation}
\widehat{D}_{j}\left(  r\right)  =\left(
\begin{array}
[c]{ccc}%
J_{j-1/2}\left(  \widetilde{\kappa}r\right)  &  & J_{j+1/2}\left(
\widetilde{\kappa}r\right) \\
&  & \\
N_{j-1/2}\left(  \widetilde{\kappa}r\right)  &  & N_{j+1/2}\left(
\widetilde{\kappa}r\right)
\end{array}
\right)  \label{D}%
\end{equation}

Then eq.(\ref{C}) can be written as follows%
\begin{equation}
C_{j}=\frac{\mathcal{J}_{j,1}}{\mathcal{J}_{j,2}}, \label{C2}%
\end{equation}
where the vector $\mathcal{J}_{j}=\left(
\begin{array}
[c]{c}%
\mathcal{J}_{j,1}\\
\\
\mathcal{J}_{j,2}%
\end{array}
\right)  $ is determined as follows%
\begin{equation}
\mathcal{J}_{j}\left(  r_{1}\right)  =\widehat{D_{j}}\left(  r_{1}\right)
\mathbf{J}_{j}\left(  r_{1}\right)  . \label{jtilda}%
\end{equation}
Similarly, introducing the transformed vector $\mathbf{H}_{j}^{\left(
\alpha\right)  }\left(  r_{2}\right)  =\widehat{D}\left(  r_{2}\right)
\mathbf{h}^{\left(  \alpha\right)  }\left(  r_{2}\right)  ,$ we can wtite the
S-matrix (\ref{S}), (\ref{F}) in a simpler form:%
\begin{equation}
S_{j}=-\frac{\mathbf{H}_{j,1}^{(2)}\mathcal{J}_{j,2}+\mathbf{H}_{j,2}%
^{(2)}\mathcal{J}_{j,1}}{\mathbf{H}_{j,1}^{(1)}\mathcal{J}_{j,2}%
+\mathbf{H}_{j,2}^{(1)}\mathcal{J}_{j,1}}. \label{S2}%
\end{equation}
This formula can be written in another form introducing the components
transposition operator $\widehat{P}=\widehat{\sigma}_{1}:$
\begin{equation}
S_{j}=-\frac{\left(  \widehat{D}_{j}\left(  r_{1}\right)  \mathbf{J}%
_{j,1}\left(  r_{1}\right)  ,\widehat{\sigma}_{1}\widehat{D}_{j}\left(
r_{2}\right)  \mathbf{h}_{j}^{\left(  2\right)  }\left(  r_{2}\right)
\right)  }{\left(  \widehat{D}_{j}\left(  r_{1}\right)  \mathbf{J}%
_{j,1}\left(  r_{1}\right)  ,\widehat{\sigma}_{1}\widehat{D}_{j}\left(
r_{2}\right)  \mathbf{h}_{j}^{\left(  1\right)  }\left(  r_{2}\right)
\right)  }, \label{S3}%
\end{equation}
where $\left(  \mathbf{a},\mathbf{b}\right)  $ is a scalar product of the
vectors $\mathbf{a}$ and $\mathbf{b}$. The formula (\ref{S3}) can be
re-written in the form:%
\begin{equation}
S_{j}=-\frac{\left(  \widehat{K_{j}}\mathbf{J}_{j}\left(  r_{1}\right)
,\mathbf{h}_{j}^{\left(  2\right)  }\left(  r_{2}\right)  \right)  }{\left(
\widehat{K_{j}}\mathbf{J}_{j}\left(  r_{1}\right)  ,\mathbf{h}_{j}^{\left(
1\right)  }\left(  r_{2}\right)  \right)  }, \label{S4}%
\end{equation}

where the matrix $\widehat{K}_{j}$ is determined as follows%
\begin{equation}
\widehat{K}_{j}=\widehat{D}_{j}^{\dagger}\left(  r_{2}\right)  \widehat
{\sigma}_{1}\widehat{D}_{j}\left(  r_{1}\right)  . \label{K}%
\end{equation}

The characteristic equation now reads
\begin{equation}
\left(  \widehat{K}_{j}\mathbf{J}_{j}\left(  r_{1}\right)  ,\mathbf{h}%
_{j}^{\left(  1\right)  }\left(  r_{2}\right)  \right)  =0 \label{char2}%
\end{equation}

\section{S- and T-matrix properties. Possible applications}

Using the relations $H_{n}^{\left(  1\right)  }\left(  z\right)  =J_{n}%
+iN_{n},$ $H_{n}^{\left(  2\right)  }=J_{n}-iN_{n},$ we can write S-matrix in
the form:%
\begin{equation}
S_{j}\left(  E\right)  =-\frac{A_{j}\left(  E\right)  +iB_{j}\left(  E\right)
}{A_{j}\left(  E\right)  -iB_{j}\left(  E\right)  }=\frac{B_{j}\left(
E\right)  +iA_{j}\left(  E\right)  }{B_{j}\left(  E\right)  -iA_{j}\left(
E\right)  }, \label{swrational}%
\end{equation}
and, therefore, it can be presented in the standard form \cite{KMLL}%
\begin{equation}
S_{j}\left(  E\right)  =\exp\left[  i2\delta_{j}\left(  E\right)  \right]  ,
\label{phase}%
\end{equation}
where the scattering phase is given by the expression
\begin{equation}
\delta_{j}\left(  E\right)  =\arctan\frac{A_{j}\left(  E\right)  }%
{B_{j}\left(  E\right)  }. \label{arctan}%
\end{equation}
Formulae (\ref{swrational}), (\ref{phase}) show once more that the scattering
matrix $S_{j}\left(  E\right)  $ is unitary on the continuum spectrum. The
functions $A_{j}\left(  E\right)  $ and $B_{j}\left(  E\right)  $ are
determined as follows:%
\begin{equation}
A=\sqrt{\frac{E+m}{E-m}}J_{j-1/2}\left(  \kappa r_{2}\right)  \left[
J_{j+1/2}\left(  \widetilde{\kappa}r_{2}\right)  +C_{j}N_{j+1/2}\left(
\widetilde{\kappa}r_{2}\right)  \right]  -\sqrt{\frac{\widetilde{E}%
+\widetilde{m}}{\widetilde{E}-\widetilde{m}}}J_{j+1/2}\left(  \kappa
r_{2}\right)  \left[  J_{j-1/2}\left(  \widetilde{\kappa}r_{2}\right)
+C_{j}N_{j-1/2}\left(  \widetilde{\kappa}r_{2}\right)  \right]  , \label{A}%
\end{equation}%
\begin{equation}
B=\sqrt{\frac{\widetilde{E}+\widetilde{m}}{\widetilde{E}-\widetilde{m}}%
}N_{j+1/2}\left(  \kappa r_{2}\right)  \left[  J_{j-1/2}\left(  \widetilde
{\kappa}r_{2}\right)  +C_{j}N_{j-1/2}\left(  \widetilde{\kappa}r_{2}\right)
\right]  -\sqrt{\frac{E+m}{E-m}}N_{j-1/2}\left(  \kappa r_{2}\right)  \left[
J_{j+1/2}\left(  \widetilde{\kappa}r_{2}\right)  +C_{j}N_{j+1/2}\left(
\widetilde{\kappa}r_{2}\right)  \right]  \label{B}%
\end{equation}
where the constant $C_{j}$ is given by eq. (\ref{C}). It is seen from
(\ref{C}), (\ref{A}), and (\ref{B}) that all $\delta_{j}\left(  E\right)  $
($j$=$\pm1/2,\pm3/2\ldots)$ vanish , when $a$ and $b$ tend to zero, i. e. in
the absence of a perturbation.

\bigskip It is easy to show that the phase is proportional to $\kappa r_{0}$
in the long-wave limit as it is necessary \cite{KMLL}, \cite{novikov} (here
$r_{0}\sim r_{1}$, $r_{2}$) (see also \cite{we2}). The scattering amplitude
$f\left(  \theta\right)  $ and transport cross-section $\Sigma_{tr}$ can be
expressed in terms of $S_{j}\left(  E\right)  $ as follow \cite{novikov}:%
\begin{equation}
f\left(  \theta\right)  =\frac{1}{i\sqrt{2\pi\kappa}}\sum_{j=\pm
1/2,\pm3/2,...}\left[  S_{j}\left(  E\right)  -1\right]  \exp\left[  i\left(
j-1/2\right)  \theta\right]  , \label{amplitude}%
\end{equation}%
\begin{equation}
\Sigma_{tr}=2/\kappa\sum_{j=\pm1/2,\pm3/2,..}\sin^{2}\left(  \delta
_{j+1}-\delta_{j}\right)  \label{crosssection}%
\end{equation}
In the vicinity of the resonance state energy, the Breit-Wigner form of the
phase is valid \cite{KMLL}:
\begin{equation}
\delta_{j}\approx\delta_{j}^{\left(  0\right)  }+\arctan\frac{\Gamma_{j}%
}{2\left(  E_{j}^{(0)}-E\right)  }, \label{breit}%
\end{equation}
where $E_{j}^{(0)}$ and $\Gamma_{j}$ are respectively the position and width
of the resonance level, $\delta_{j}^{\left(  0\right)  }$ is the
slowly-varying potential scattering phase. The presented above formulae can be
used in order to calculate the Boltzmann conductivity \cite{sarma}:
\begin{equation}
\sigma=\left(  \frac{e^{2}}{2\pi\hslash}\right)  \frac{2E_{F}}{\hslash}%
\tau_{tr}, \label{conductivity.}%
\end{equation}
where the transport relaxation time equals%

\begin{equation}
1/\tau_{tr}=N_{i}v_{F}\Sigma_{tr}. \label{relaxtime}%
\end{equation}
Here $N_{i}$ is the impurities areal density, $E_{F}=v_{F}\kappa_{F}$. The
above equations transform the scattering data into the correspondent
dependence of the Boltzmann conductivity. Thus characteristic features of the
scattering data determine a behaviour of the electric conductivity. Proper
numeric calculations will be presented elsewhere.

On the other hand, the derived here formulae for scattering data can be used
to obtain an approximate density of states and other observables in lower
order to the impurities density. The exact explicit formula for one-impurity
S-matrix obtained here, which allows us to calculate the scattering
amplitudes, bound and resonance states, determines also the on-shell
one-impurity T-matrix \cite{baz}:%

\begin{equation}
T_{j}^{on}\left(  E\right)  =\left(  1/i\right)  \left[  S_{j}\left(
E\right)  -1\right]  \label{TS}%
\end{equation}
Corresponding off-shell T-matrix $\widehat{T}^{off}\left(  \mathbf{k,k}%
^{\prime}\mathbf{,}E\right)  $ (here $\mathbf{k}$ generically does not lie on
the mass shell, i. e. $\mathbf{k}^{2}+m^{2}$ can be not equal to $E^{2}$) can
be written in the form \cite{newton}:
\begin{equation}
T^{off}\left(  \mathbf{k,k}^{\prime},E\right)  =\left(  \Psi_{0}%
(\mathbf{k})\widehat{T}^{off}\left(  E\right)  \Psi_{0}\left(  \mathbf{k}%
^{\prime}\right)  \right)  . \label{Toff}%
\end{equation}
where the transition operator $\widehat{T}^{off}\left(  E\right)  $ is
determined by the operator equation \cite{baz}:%
\begin{equation}
\widehat{T}^{off}=\widehat{U}-\widehat{U}\widehat{G}_{0}\widehat{T}^{off}.
\label{operatorequation}%
\end{equation}
Here $\widehat{G}_{0}=\widehat{H}_{0}^{-1}$ is the operator inverse to the
free Hamiltonian. All components of the product in eq. (\ref{Toff}) are
determined at different energies: $E,$ $E\left(  \mathbf{k}\right)  $ and
$E\left(  \mathbf{k}^{\prime}\right)  $. Let us write the Lippmann-Schwinger
equation (\ref{operatorequation}) for off-shell one-impurity T-matrix in the
matrix form:%

\begin{equation}
T^{^{off}}\left(  \mathbf{k,k}^{\prime},E\right)  =U\left(  \mathbf{k-k}%
^{\prime}\right)  -\int d^{2}qU\left(  \mathbf{k-q}\right)  \frac{\gamma_{\mu
}q_{\mu}+m-\gamma_{0}E}{\mathbf{q}^{2}+m^{2}-E^{2}+i0}T^{^{off}}\left(
\mathbf{q,k}^{\prime},E\right)  , \label{lippmann}%
\end{equation}
where the Fourier transform of the potential is determined as follows%

\begin{equation}
U\left(  \mathbf{\mathbf{k-}k}^{\prime}\right)  =\int d^{2}r\exp\left[
-i\left(  \mathbf{k-k}^{\prime}\right)  \mathbf{r}\right]  U(r)=2\pi\sum
_{m}\epsilon_{m}\cos m\theta\int_{0}^{\infty}drrJ_{m}\left(  kr\right)
U\left(  r\right)  J_{m}\left(  k^{\prime}r\right)  , \label{fourier}%
\end{equation}

\begin{equation}
\epsilon_{m}=\left\{
\begin{array}
[c]{c}%
1\text{ \ \ if }m=0,\\
2\text{ \ \ if }m\neq0.
\end{array}
\right.  , \label{epsylon}%
\end{equation}
$\gamma_{\mu}=\beta\alpha_{\mu}\mathbf{,}$ $\gamma_{0}=\beta$, $\theta$ is the
angle between the vectors $\mathbf{k}$ and $\mathbf{k}^{\prime}.$

Now we consider a question: how the obtained above formalae for scattering
data can be used for calculation of observables in the monolayer graphene. It
is seen from (\ref{fourier}) that in the case of the short-range perturbation,
$U\left(  \mathbf{q}\right)  $ is a slowly-varying function. Then equation
(\ref{lippmann}) shows that the off-shell T-matrix is a slowly-varying
function of $\mathbf{k}$ and $\mathbf{k}^{\prime}$ as well. At the same time,
it can be a sharp function of energy $E$ in the vicinity of a resonance.
Therefore, in the case of a narrow resonance, the most important information
on the off-shell transition matrix is given by the on-shell matrix, which is
determined by the S-matrix (see (\ref{TS})).%
\begin{equation}
T^{^{off}}\left(  \mathbf{k,k}^{\prime},E\right)  \simeq T^{^{on}}\left(
\mathbf{k,k}^{\prime},E\right)  =2i\sum_{j}\epsilon_{j-1/2}\cos\left[  \left(
j-1/2\right)  \theta\right]  \left\{  S_{j}\left(  E\right)  -1\right\}  .
\label{TS2}%
\end{equation}
Using the derived above explicit formula for S-matrix (\ref{S4}) (or
(\ref{S})), an approximate formula for the off-shell one-impurity T-matrix in
the momentum representation can be obtained. Substituting the partial-wave
projection of the S-matrix (\ref{phase}) into equation (\ref{TS2}), we can
obtain an expression for the T-matrix, which can be used for approximate
calculation of such observables as the denstity of states, for instance. The
scattering phases $\delta_{j}\left(  E\right)  $ are completely determined by
the formulae (\ref{A}), (\ref{B}) and (\ref{arctan}) obtained in this paper .
The electronic density of states and other observables for systems with
non-overlapping impurity potentials can be calculated on the base of Lloyd's
formula (see, for instance, \cite{Yakibchuk}). The simplest approximation can
be obtained in the lower-order approximation to the mass operator $M\left(
\mathbf{k,E}\right)  $ in the impurities density $n$ using the mean T-matrix
method for a random distribution of impurities \cite{meanT}:
\begin{equation}
M\left(  \mathbf{k,E}\right)  =n\left\langle \mathbf{k}\right\vert
T^{off}\left(  E\right)  \left\vert \mathbf{k}\right\rangle \label{mass}%
\end{equation}

Thus, having the above formulae for the scattering data, we can calculate the
averaged over impurities distribution Green functions, density of states and
the optical absorption coefficient. Results of such calculations will be
published elsewhere.

\section{Conclusion}

We considered the electron scattering problem in the monolayer graphene with
short-range impurities. Characteristic for the electronic two-band theory band
asymmetry of the potential is equivalently described by the scalar potential
and the mass (gap) local perturbation. The crystal perturbation by a single
impurity is modelled by the annular well. Exact explicit formulae for a
single-impurity S-matrix and other scattering data have been obtained and
analyzed for the short-range perturbation modelled with a use of the annular
well with the band-asymmetric potential. The characteistic equation for bound
and resonance states is derived for the case. This equation was shown to have
a form of 2-vectors orthogonality condition. Possible application of these
results to description of kinetical and optical properties of graphene is
discussed. A procedure of approximate calculation of observables based on a
substiution of the off-shell T-matrix by the one-impurity on-shell T-matrix,
which can be expressed in terms of an exact S-matrix in the vicinity of a
sharp resonance is suggested. The obtained in this paper explicit formulae for
S-matrix can be used in this procedure. As an elementary example of this
approach, we considered an approximate calculation of the mass operator in
graphene with non-overlapping defects.

\end{document}